\documentclass[twoside,a4paper]{article}
\usepackage{fleqn,epsf,graphicx}
\usepackage{authblk}
\usepackage{hyperref}
\usepackage{xcolor}
\hypersetup{
  colorlinks   = true, %Colours links instead of ugly boxes
  urlcolor     = black, %Colour for external hyperlinks
  linkcolor    = black, %Colour of internal links
  citecolor   = black %Colour of citations
}
\usepackage{cite}
\usepackage{tikz}
\usetikzlibrary{arrows,calc, shapes, decorations.pathreplacing}
\begin{document}
\sloppy
\title{Sampling FEE and Trigger-less DAQ for the J-PET Scanner%
\thanks{Presented at Jagiellonian Symposium on Fundamental and Applied Subatomic Physics}%
}

\author{G.~Korcyl$^{a}$, D. Alfs$^{a}$, T.~Bednarski$^{a}$, P.~Bia\l as$^{a}$, E.~Czerwi\'nski$^{a}$, K.~Dulski$^{a}$, A.~Gajos$^{a}$, B.~G\l owacz$^{b}$, B. Jasi\'nska$^{b}$, D.~Kami\'nska$^{a}$, \L .~Kap\l on$^{a,d}$, P.~Kowalski$^{f}$, T.~Kozik$^{a}$, W.~Krzemie\'n$^{c}$, E.~Kubicz$^{a}$, M.~Mohammed$^{a}$, Sz.~Nied\'zwiecki$^{a}$,  M.~Pa\l ka$^{a}$, M.~Pawlik-Nied\'zwiecka$^{a}$, L.~Raczy\'nski$^{f}$, Z.~Rudy$^{a}$, O.~Rundel$^{a}$, N.G.~Sharma$^{a}$, M.~Silarski$^{a}$, A.~S\l omski$^{a}$, K.~Sto\l a$^{a}$, A.~Strzelecki$^{a}$, A.~Wieczorek$^{a,d}$, W.~Wi\'slicki$^{f}$, B. K.~Zgardzi\'nska$^{b}$, M.~Zieli\'nski$^{a}$, P.~Moskal$^{a}$}

\affil{
       
       $^{a}$Faculty of Physics, Astronomy and Applied Computer Science, Jagiellonian University, 30-348 Cracow, Poland\\
       $^{b}$Department of Nuclear Methods, Institute of Physics, Maria Curie-Sk\l odowska University, 20-031 Lublin, Poland\\
       $^{c}$High Energy Physics Division, National Center for Nuclear Research, 05-400 Otwock-\'Swierk, Poland\\
       $^{d}$Institute of Metallurgy and Materials Science of Polish Academy of Sciences, 30-059 Cracow, Poland\\
       $^{e}$Faculty of Chemistry, Jagiellonian University, 30-060 Cracow, Poland\\
       $^{f}$\'Swierk Computing Center, National Center for Nuclear Research, 05-400 Otwock-\'Swierk, Poland\\
     }

\maketitle
\begin{abstract}
In this paper, we present a complete Data Acquisition System (DAQ) together with the readout mechanisms for the J-PET tomography scanner. In general detector readout chain is constructed out of Front-End Electronics (FEE), measurement devices like Time-to-Digital or Analog-to-Digital Converters (TDCs or ADCs), data collectors and storage. We have developed a system capable for maintaining continuous readout of digitized data without preliminary selection. Such operation mode results in up to 8 Gbps data stream, therefore it is required to introduce a dedicated module for online event building and feature extraction. The Central Controller Module, equipped with Xilinx Zynq SoC and 16 optical transceivers serves as such true real time computing facility. Our solution for the continuous data recording (trigger-less) is a novel approach in such detector systems and assures that most of the information is preserved on the storage for further, high-level processing. Signal discrimination applies an unique method of using LVDS buffers located in the FPGA fabric.
\end{abstract}
  
\section{Introduction}
The prototype of TOF-PET scanner, constructed by the J-PET collaboration \cite{mpatent1, mpatent2, nim1, nim2, nim3, nim4, nim5, nim6, nim7, nim8}, consists of 192 plastic scintillators, each equipped with photomultipliers (PMTs) on both ends. This results in 384 analog channels that have to be processed by the Data Acquisition System (DAQ). The J-PET prototype will be used for investigations in the field of medical imaging~\cite{nim3,x2}, nano-biology~\cite{x3}, material science~\cite{x4,x5} and for testing of fundamental symmetries in physics~\cite{x6,x7}. In this paper, we present a complete solution for the DAQ system together with the readout mechanisms. Detector readout chain is constructed out of Front-End Electronics (FEE), measurement devices like Time-to-Digital or Analog-to-Digital Converters (TDCs or ADCs), data collectors and storage. Most of the PET scanners include coincidence units or multi-level trigger logic in order to discard in the real-time data classified as background noise \cite{US7091489B2,US8164063B2}. Such a unit can only execute low-level selection algorithms in order to fulfill real-time regime. Applying low-level rejection filters can result in the loss of fraction of valuable data. On the other hand, more complex algorithms might introduce longer dead-time of the DAQ system, resulting in reduced rate of registered events. Our solution for the continuous data recording (trigger-less) is a novel approach in such detector systems and assures that most of the information is preserved on the storage for further, high-level processing. The core of the presented system is based on Trigger~Readout~Board~v3~(TRBv3) platform \cite{trb3_greg, trb3_neiser}, developed for and widely used in high energy physics experiments \cite{hades}.

\section{System Overview}
\begin{figure}[htb]
\centerline{%
\includegraphics[width=12.5cm]{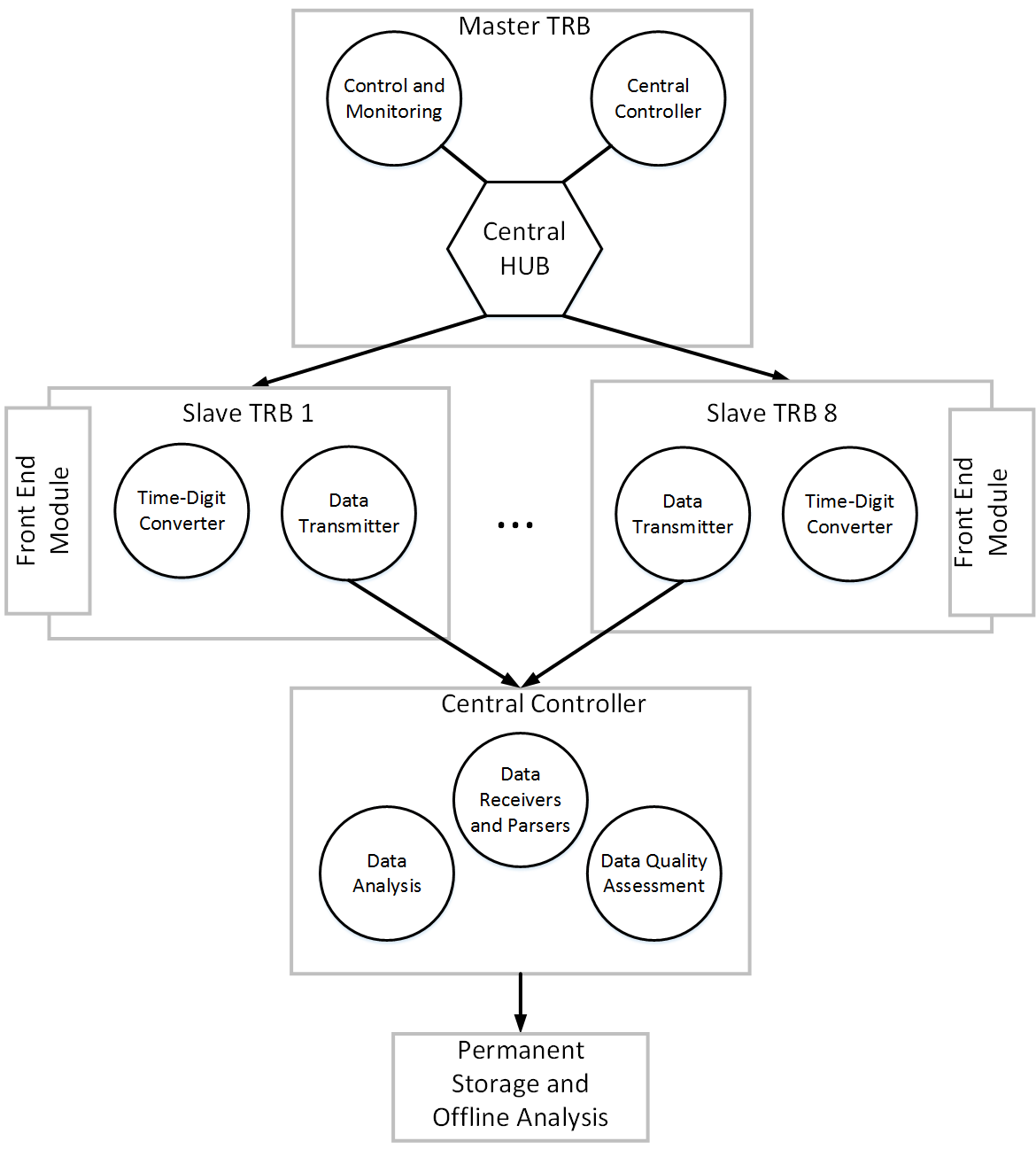}}
\caption{Schematic view of the JPET Data Acquisition Setup. The readout process is controlled by the Master TRBv3 module, whereas Slave TRBv3 modules perform digitization of detector signals coming through Front End Modules. The readout data is streamed to the Central Controller Module and then further to permanent storage.  }
\label{Fig:F2H}
\end{figure}

The main element of the J-PET DAQ system is the collection of TRBv3 modules. Those are high-performance and FPGA-based (Field Programmable Gate Array), therefore reconfigurable, TDC readout boards. Each module is equipped with five Lattice ECP3M devices. The central one serves as the controller and local data collector, while the remaining four can be configured with configware providing various functionality. For the precise time measurement, a design providing 48 input channels and time resolution of 12 ps has been developed \cite{fpga_tdc}. Each input channel has rising and falling signal edge detection and a buffer for up to 54 complete signals storage between two consecutive readouts.
One TRBv3 module is called Master and controls the readout procedure and synchronization of all the other modules (Slaves) \ref{Fig:F2H}. Each TRBv3 board has an individual Gigabit Ethernet (GbE) link for transmitting collected measurement data out of the system for storage with the use of standard and cheap network facility.
An additional module, called Central Controller Module (CCM) has been developed in order to provide efficient, online data processing. The module has sixteen GbE links as inputs from the Slave modules and a Xilinx Zynq-7045 as the processing unit. Data packages sent from the Slaves can be directed through the CCM, which can perform online analysis, histogramming and data quality assessment.
The data is saved on the Event Building (EB) machines. Those are server class, multiprocessor PCs running software for collecting data fragments from the network and reassembling them into complete data units called Events, representing the state of the entire detector in a particular period of time. Such files are taken as the input into the analysis and image reconstruction algorithms \cite{krzemien,krzemien2}.

\subsection{Front-End Electronics}
Signal shaping, amplification and discrimination are the main functions of the Front-End Electronics \cite{palkapatent}. A dedicated electronic module has been developed as the mezzanine Add-on for the TRBv3 boards. The module is plugged directly into the connector which provides power, data and control lines.
The signal discrimination applies a novel concept of misusing (using in a non-standard way) the LVDS buffer inside the FPGA device, which is configured with time measurement logic. An LVDS buffer has two inputs: positive and negative. The output state of the buffer changes the logic state in case the voltage levels on the inputs are crossed. One can apply on one of the inputs a threshold level and the analog signal on the second input. At the moment when the analog signal crosses the threshold level, the buffer will generate a logic pulse, which being already inside the FPGA, can be directly inserted into the tapped-delay chain for precise time measurement.
The input signals are amplified and split into four paths, each having an individual threshold level. It is a realization of the multi-level thresholding concept as a measure for reducing the time-walk effect and therefore achieving better timing resolution \cite{palka}.
The TDC design allows for measurement of 48 input channels, hence the FEE module has 12 inputs from the photomultipliers.

\subsection{Readout Procedure}
The setup for the J-PET prototype consists of eight TRBv3 boards as Slave modules and one Master TRBv3. The readout process in the system is controlled by the Master module, which has the Central Trigger System (CTS) functionality implemented in the central FPGA device. For the continuous type of readout \cite{korcylpatent}, the CTS sends a periodic Readout Request message to all the Slave modules at the fixed rate of 50 kHz. The Slaves record input signals and store the data in buffers until Readout Request message arrives. At the moment the message arrives, the buffers are cleared and the data is encapsulated into UDP packets and sent over Gigabit Ethernet network to the Event Building machines. A single UDP packet contains measured time samples recorded by one Slave modules during a period of 20~$\mu$s. The readout rate and buffers sizes (up to 54 signals per channel) are adjusted accordingly in order not to get overflowed and therefore to achieve maximum signals acceptance.
Upon receiving a Readout Request message, each module constructs a UDP packet, tagged with the Readout Request message sequence number and module ID. Those two values are necessary for the Event Building machines in order to correctly combine packets from the same measurement but different sources into one, single data unit.

\subsection{Central Controller Module}
The described readout type results in a significant constant stream of data that has to be processed. The maximum throughput that can be achieved reaches 8 Gbps. This amount of data can be efficiently distributed for storage over a number of Event Building machines running in parallel and connected with 10G Ethernet network.
However the high event rate and the distributed data storage makes the designed system not suitable for online analysis. To overcome this drawback and additional module called Central Controller Module (CCM) has been designed and developed. The Central Controller Module provides a computing facility for online analysis and data quality assessment. The board features a Xilinx Zynq-7045 FPGA device, which is a hybrid of FPGA resources and an ARM processor. The architecture of FPGA devices allows for natural parallelism for processing multiple data streams, while the standard processor provides a convenient access to the results.
The module can be used as the “board-in-the-middle”, meaning it can receive the packet streams from Slave modules, perform analysis and forward the original packets further, to Event Building machines. TRBv3 data format parsers and feature extraction algorithms have been implemented as the foundation for higher-level data analysis.

\section{Summary and Outlook}
A solution for the complete Data Acquisition System for J-PET prototype has been developed and is under evaluation. It consists of two novel approaches: in-FPGA signal discrimination and continuous type of the readout \cite{korcyl}. Those two techniques allow for collection of timing data, measured with the high resolution. At present, lack of real-time data selection and filtering mechanisms in favor of trigger-less readout results in significant amount of data for processing but reduces the risk of discarding valuable measurements. Ongoing work focuses on hardware evaluation and development of the online algorithms for the Central Controller Module. 

\section{Acknowledgements}
We acknowledge technical and administrative support by  A. Heczko, M. Kajetanowicz, W. Migda\l, and the financial support by The Polish National Center for Development and Research through grant No. INNOTECH-K1/IN1/64/159174/NCBR/12, The Foundation for Polish Science through MPD program and the EU, MSHE Grant No. POIG.02.03.00-161 00-013/09.

\end{document}